\documentclass[UTF-8,preprint,showpacs,preprintnumbers,superscriptaddress]{revtex4-1}
\usepackage{multirow}
\usepackage{diagbox}
\usepackage{booktabs}
\usepackage{array}
\usepackage{natbib}
\usepackage{amsmath,amssymb,graphicx,bm,subfigure,float,siunitx,color}
\usepackage{bm,subfigure,float,siunitx,graphicx}
\usepackage[utf8]{inputenc}
\usepackage[figuresright]{rotating}
\usepackage{color}
\usepackage{epstopdf}
\usepackage{mathrsfs}

\begin{document}

\title{Regular hairy black holes through gravitational decoupling method}

\author{Yaobin Hua}
\affiliation{College of Physics Science and Technology, Hebei University, Baoding 071002, China}

\author{Zhenglong Ban}
\affiliation{College of Physics Science and Technology, Hebei University, Baoding 071002, China}

\author{Tian-You, Ren}
\affiliation{College of Physics Science and Technology, Hebei University, Baoding 071002, China}

\author{Jia-Jun, Yin}
\affiliation{College of Physics Science and Technology, Hebei University, Baoding 071002, China}

\author{Rong-Jia Yang \footnote{Corresponding author}}
\email{yangrongjia@tsinghua.org.cn}
\affiliation{College of Physics Science and Technology, Hebei University, Baoding 071002, China}
\affiliation{Hebei Key Lab of Optic-Electronic Information and Materials, Hebei University, Baoding 071002, China}
\affiliation{National-Local Joint Engineering Laboratory of New Energy Photoelectric Devices, Hebei University, Baoding 071002, China}
\affiliation{Key Laboratory of High-pricision Computation and Application of Quantum Field Theory of Hebei Province, Hebei University, Baoding 071002, China}

\begin{abstract}
Within a framework requiring a well-defined event horizon and matter obeying the weak energy condition, we employ gravitational decoupling method to construct non-singular hairy black holes: spherically or axially symmetric. These solutions arise from a deformation of the Minkowski vacuum, where the maximum deformation can yield the Schwarzschild metric for the static case, and the Kerr geometry for the stationary case, respectively.
\end{abstract}

\maketitle

\section{introduction}
The possible conditions for circumventing the no-hair conjecture have been extensively studied in different scenarios \cite{Martinez:2004nb,Sotiriou:2011dz,babichev2014dressing,Antoniou:2017acq,Antoniou:2017hxj,Sotiriou:2013qea}. One promising approach involves populating the static vacuum of General Relativity (GR) with an additional source, often described by a scalar field \cite{Herdeiro:2015waa}, which may have a fundamental origin. A primary motivation for this is to eliminate the singularities predicted to form at the end of gravitational collapse in GR. Although the Cosmic Censorship Conjecture (CCC) posits that such singularities are always hidden within event horizons \cite{Penrose:1969pc,hawking2023large}, their very existence underscores a fundamental limitation of the theory.

The pursuit of singularity resolution through hairy solutions has led to numerous proposals for regular black holes in recent years. Among these, models based on non-linear electrodynamics offer a relatively straightforward framework for sourcing singularity-free geometries \cite{Salazar:1987ap,Ayon-Beato:1998hmi,Bronnikov:2000vy,Dymnikova:2004zc,Balart:2014cga,Toshmatov:2014nya,Fan:2016hvf}. However, such classically regular solutions typically involve a Cauchy horizon—a null hypersurface that compromises deterministic predictability \cite{Poisson:1989zz,Poisson:1990eh} and introduces a host of theoretical difficulties \cite{Bonanno:2020fgp,carballo2022regular,Franzin:2022wai,Bonanno:2022jjp,Ovalle:2023ref}. A promising path toward evading these issues is to describe matter in the most general terms possible, thereby ensuring a flexible scenario governed by minimal assumptions.

This work adopts precisely this approach by filling the Schwarzschild vacuum with a general static, spherically symmetric source $\theta_{\mu\nu}$, a "tensor vacuum". This construction follows directly from the gravitational decoupling (GD) method \cite{Ovalle:2017fgl,Ovalle:2018gic}, a powerful technique for generating hairy black holes in both spherically symmetric \cite{Ovalle:2018umz,Ovalle:2020kpd} and axially symmetric geometries \cite{Contreras:2021yxe,Islam:2021dyk,daRocha:2020gee,Ovalle:2021jzf,afrin2021parameter,Ramos:2021jta,Meert:2021khi,Mahapatra:2022xea,Cavalcanti:2022cga,Omwoyo:2021uah,Avalos:2023jeh,Avalos:2023ywb}. A key advantage of this scheme is its capacity to incorporate minimal physical requirements while preserving the asymptotic vacuum behavior. Our objective is thus to identify regular black hole solutions that satisfy the weak energy condition (WEC) in both static and rotating regimes. As a result, we have successfully constructed an asymptotically flat, regular black hole and systematically elucidated the influence of the relevant parameters on its properties.

The structure of this paper is as follows:
In Sec. \MakeUppercase{\romannumeral2}, we briefly introduce the GD scheme and show the decoupling of two gravitational sources in the spherically symmetric case. In Sec. \MakeUppercase{\romannumeral3}, we apply the GD method to generate spherically symmetric regular hairy BHs which satisfy the WEC. In Sec. \MakeUppercase{\romannumeral4}, we derive the axially symmetric version of the regular hairy BH. Finally in Sec. \MakeUppercase{\romannumeral6}, we summarize our conclusions.  

\section{gravitational decoupling}
We first briefly introduce the GD scheme in spherically symmetric gravitational systems, for details see \cite{Ovalle:2020kpd}. We begin with the Einstein-Hilbert action
\begin{align}
\label{EHA}
\mathcal{S}=\int \bigg[\frac{R}{2\kappa}+\mathcal{L}_{M}+\mathcal{L}_{\Theta} \bigg]\sqrt{-g}\ d^{4}x,
\end{align}
where $\kappa=8\,\pi\,G$ with unit $c=1$, $R$ is the Ricci scalar, $\mathcal{L}_{M}$ represents the standard matter fields, and
$\mathcal{L}_{\Theta}$ is the Lagrangian density describing matter or being related to the new gravitational sectors beyond GR. For these two sources, the energy-momentum tensor is generally defined as
\begin{align}
T_{\mu\nu}&=-\frac{2}{\sqrt{-g}}\frac{\delta\, \big(\sqrt{-g}\,\mathcal{L}_{M}\big)}{\delta \,g^{\mu\nu}}=g_{\mu\nu}\,\mathcal{L}_{M}-2\,\frac{\delta\,\mathcal{L}_{M}}{\delta \,g^{\mu\nu}} \label{Tuv},   \\
\theta_{\mu\nu}&=-\frac{2}{\sqrt{-g}}\frac{\delta\, \big(\sqrt{-g}\,\mathcal{L}_{\Theta}\big)}{\delta \,g^{\mu\nu}}=g_{\mu\nu}\,\mathcal{L}_{\Theta}-2\,\frac{\delta\,\mathcal{L}_{\Theta}}{\delta \,g^{\mu\nu}}\label{thetauv}.
\end{align}
From the action \eqref{EHA}, one can derive the Einstein field equations
\begin{align}
\label{Guv}
G_{\mu\nu}=R_{\mu\nu}-\frac{1}{2}R\,g_{\mu\nu}=\kappa \,\tilde{T}_{\mu\nu},
\end{align}
where $\tilde{T}_{\mu\nu}$ represents the total energy-momentum tensor: $\tilde{T}_{\mu\nu}=T_{\mu\nu}+\theta_{\mu\nu}$, which must be covariantly conserved: $\nabla_{\mu} \, \tilde{T}^{\mu\nu}=0$, according to the second Bianchi identity.

For a static and spherically symmetric system, the metric can be writen as
\begin{align}
\label{1g}
ds^{2}=-e^{A(r)}dt^{2}+e^{B(r)}dr^{2}+r^{2}\Big(d\theta^{2}+\sin^{2}\theta d\phi^{2}\Big),
\end{align}
where $A=A(r)$ and $B=B(r)$ are a function of the radial coordinates $r$. 
Then, the Einstein field equations Eq. \eqref{Guv} give
\begin{align}
G_{0}^{\ 0}&=\kappa \,\tilde{T}_{0}^{\ 0}=\kappa \,\big(T_{0}^{\ 0}+\theta_{0}^{\ 0}\big)=-\frac{1}{r^{2}}+e^{-B}\Bigg(\frac{1}{r^{2}}-\frac{B'}{r}\Bigg),   \label{G00} 
\end{align}
\begin{align}
G_{1}^{\ 1}&=\kappa \,\tilde{T}_{1}^{\ 1}=\kappa \,\big(T_{1}^{\ 1}+\theta_{1}^{\ 1}\big)=-\frac{1}{r^{2}}+e^{-B}\Bigg(\frac{1}{r^{2}}+\frac{A'}{r}\Bigg),    \label{G11} 
\end{align}
\begin{align}
G_{2}^{\ 2}&=\kappa \,\tilde{T}_{2}^{\ 2}=\kappa \,\big(T_{2}^{\ 2}+\theta_{2}^{\ 2}\big)=\frac{e^{-B}}{4}\Bigg(2A''+A'^{2}-A'B'+2\,\frac{A'-B'}{r}\Bigg),   \label{G22} 
\end{align}
\begin{align}
G_{3}^{\ 3}&=\kappa \,\tilde{T}_{3}^{\ 3}=\kappa \,\big(T_{3}^{\ 3}+\theta_{3}^{\ 3}\big)=\frac{e^{-B}}{4}\Bigg(2A''+A'^{2}-A'B'+2\,\frac{A'-B'}{r}\Bigg),   \label{G33}
\end{align} 
where $f'\equiv\partial_{r}f$. Since the metric is spherically symmetric, it is easy to see that $G_{2}^{\ 2}=G_{3}^{\ 3}$. Next, we identify an effective energy density $\tilde{\epsilon}$, an effective radial pressure $\tilde{p_{r}}$, and an effective tangential pressure $\tilde{p_{t}}$, respectively, as
\begin{align}
\tilde{\epsilon}&\equiv \epsilon+\mathcal{E}=-T_{0}^{\ 0}-\theta_{0}^{\ 0},                \label{epsilon}  
\end{align}
\begin{align}
\tilde{p}_{r}&\equiv p_{r}+\mathcal{P}_{r}=T_{1}^{\ 1}+\theta_{1}^{\ 1},                   \label{pr}      
\end{align}
\begin{align}
\tilde{p}_{t}&\equiv p_{\theta}+\mathcal{P}_{\theta}=T_{2}^{\ 2}+\theta_{2}^{\ 2},    \label{pt}
\end{align}
where $\epsilon$, $p_{r}$, and $p_{\theta}$ are related to $T_{\mu}^{\ \nu}$; while $\mathcal{E}$, $\mathcal{P}_{r}$, $\mathcal{P}_{\theta}$ are related to $\theta_{\mu}^{\ \nu}$. Then, we have $T_{\mu}^{\ \nu}={\rm diag}\big[-\epsilon,p_{r},~p_{\theta},~p_{\theta}\big]$, and $\theta_{\mu}^{\ \nu}={\rm diag}\big[-\mathcal{E},~\mathcal{P}_{r},~\mathcal{P}_{\theta},~\mathcal{P}_{\theta}\big]$. In general, Eqs. \eqref{G00}-\eqref{G22} describe an anisotropic fluid, and $\Pi\equiv\tilde{p}_{t}-\tilde{p}_{r}$ is not zero.

Now considering a solution for Eq. \eqref{Guv} generated by the seed source $T_{\mu\nu}$ alone, which can be written as
\begin{align}
\label{2g}
ds^{2}=-e^{D(r)}dt^{2}+e^{E(r)}dr^{2}+r^{2}\Big(d\theta^{2}+\sin^{2}\theta d\phi^{2}\Big),
\end{align}
where 
\begin{align}
\label{eFm}
e^{-E(r)}\equiv1+\frac{\kappa}{r}\int_{0}^{r} x^{2}\,T_{0}^{\ 0}\,dx=1-\frac{2\,m(r)}{r},
\end{align}
which is the standard expression in GR for $m(r)$ being the Misner-Sharp mass $m$. After adding the source $\theta_{\mu\nu}$, we have the GD of the metric \eqref{2g} as
\begin{align}
D(r)&\longrightarrow A(r)=D(r)+\alpha\,g(r),                      \label{jiantou1}
\end{align}
\begin{align}
e^{-E(r)}&\longrightarrow e^{-B(r)}=e^{-E(r)}+\alpha\,f(r),       \label{jiantou2}
\end{align}
where $g$ and $f$ are the geometric deformations for the temporal and  radial components of the metric, respectively. These geometric deformations are parameterized by $\alpha$. Note that Eqs. \eqref{jiantou1}-\eqref{jiantou2} do not represent a coordinate transformation. According to Eqs. \eqref{jiantou1}-\eqref{jiantou2}, the Einstein field Eqs. \eqref{G00}-\eqref{G22} can be divided into two sets:

The first set sourced by the energy-momentum tensor $T_{\mu\nu}$ is given by the standard Einstein field equations for the metric \eqref{2g}  
\begin{align}
\kappa \, \epsilon&=-\kappa \, T_{0}^{\ 0}=\frac{1}{r^{2}}-e^{-E}\Bigg(\frac{1}{r^{2}}-\frac{E'}{r}\Bigg) ,                                     \label{GD11} 
\end{align}
\begin{align}
\kappa \, p_{r}&=\kappa \, T_{1}^{\ 1}=-\frac{1}{r^{2}}+e^{-E}\Bigg(\frac{1}{r^{2}}+\frac{D'}{r}\Bigg),                                           \label{GD12} 
\end{align}
\begin{align}
\kappa \, p_{\theta}&=\kappa \,T_{2}^{\ 2}=\frac{e^{-E}}{4}\Bigg(2D''+D'^{2}-D'E'+2\,\frac{D'-E'}{r}\Bigg).                  \label{GD13} 
\end{align}
The second set including the source $\theta_{\mu\nu}$ is
\begin{align}
\kappa \, \mathcal{E}&=-\kappa \, \theta_{0}^{\ 0}=-\frac{\alpha\,f}{r^{2}}--\frac{\alpha\,f'}{r},             \label{GD21} 
\end{align}
\begin{align}
\kappa \, \mathcal{P}_{r}&=\kappa \, \theta_{1}^{\ 1}=\alpha\,f\Bigg(\frac{1}{r^{2}}+\frac{A'}{r}\Bigg)+\alpha\,X_{1},            \label{GD22} 
\end{align}
\begin{align}
\kappa \, \mathcal{P}_{\theta}&=\kappa \,\theta_{2}^{\ 2}=\frac{\alpha\,f}{4}\Bigg(2A''+A'^{2}+2\,\frac{A'}{r}\Bigg)+\frac{\alpha\,f'}{4}\Bigg(A'+\frac{2}{r}\Bigg)+\alpha\,X_{2}, \label{GD23} 
\end{align}
with
\begin{align}
X_{1} &=\frac{e^{-E}\,g'}{r},                                                  \label{X1} 
\end{align}
\begin{align}
4X_{2} &=e^{-E}\bigg(2\,g''+\alpha\,g'^{2}+\frac{2\,g'}{r}+2\,g'D'-E'g' \bigg).      \label{X2}
\end{align}
Obviously if the geometric deformations vanish, $f=g=0$, the tensor $\theta_{\mu\nu}$ vanishes.

Thinking of the conservation of the total energy-momentum tensor $\tilde{T}_{\mu\nu}$, we are leaded to
\begin{align}
\label{shouheng8}
\nabla_{\mu}\,T^{\ \mu}_{\nu}=\frac{\alpha\,g'}{2}\big(\epsilon+p_{r}\big)=-\nabla_{\mu}\,\theta^{\ \mu}_{\nu}.
\end{align}
Obviously, Eq. \eqref{shouheng8} indicates the energy exchange between gravitational systems, which are described by Eqs. \eqref{GD11}-\eqref{GD13} and Eqs. \eqref{GD21}-\eqref{GD23}, respectively.
Furthermore, when there is no temporal deformation $(g=0)$ or for the Kerr-Schild spacetime $(\epsilon+p_{r}=0)$, the interaction will be purely related to gravity without involving energy exchange. This result is an exact solution and there is no requirement for any perturbative expansion in $g$ and $f$ \cite{Ovalle:2020fuo}.

\section{Hairy black holes}
Now let's to consider hairy deformations of spherically symmetric BHs in GR. We begin with the Schwarzschild metric
\begin{align}
\label{seedSsolution}
e^{D}=e^{-E}=1-\frac{2M}{r},
\end{align}
which is the solution to Eqs. \eqref{GD11}-\eqref{GD13} with zero energy-momentum tensor. We take this metric as the seed geometry and search for a matter Lagrangian \( \mathcal{L}_\Theta \) corresponding to the energy–momentum tensor \( \theta_{\mu\nu} \) that induces the GD functions \( f \) and \( g \) in  Eqs. \eqref{GD21}-\eqref{GD23} to remove the singularity of the seed metric at \( r = 0 \). Since the system consists of three equations with five unknowns: \( f \), \( g \), \( E \), \( P_r \), and \( P_\theta \), We are free to choose additional conditions.

\subsection{Horizon structure}
To ensure a well-defined horizon structure for the BHs, one requires $e^{A(r_{h})}=e^{-B(r_{h})}=0$, so that both a Killing horizon ($e^{A}=0$) and a causal horizon ($e^{-B}=0$) will be satisfied simultaneously when $r=r_{h}$. For this feature we have a sufficient condition $e^{A}=e^{-B}$. This condition and the Einstein field Eqs. \eqref{G00}-\eqref{G11} lead to the equation of state $-\tilde{\epsilon}=\tilde{p_{r}}$ satisfied by the source. Obviously, we have from $T_{\mu\nu}=0$: $-\mathcal{E}=\mathcal{P}_{r}$, which indicates that for a positive energy density, the radial pressure must be negative. Combining this equation and Eqs. \eqref{shouheng8}, results is
\begin{align}
\label{Prprime}
\mathcal{P}_{r}'=\frac{2}{r}\big(\mathcal{P}_{\theta}-\mathcal{P}_{r}\big)
\end{align}
This hydrostatic equilibrium equation governs the state of the source $\theta_{\mu\nu}$, thereby preventing its collapse into the central singularity of the seed Schwarzschild metric.

Using the seed Schwarzschild solution \eqref{seedSsolution} in Eqs. \eqref{jiantou1}-\eqref{jiantou2} and the condition $e^{A}=e^{-B}$, we find
\begin{align}
\label{alphaf}
\alpha\,f=\bigg(1-\frac{2M}{r}\bigg)\bigg(e^{\alpha\,g}-1\bigg).
\end{align}
Therefore, the metric \eqref{1g} can be rewritten as
\begin{align}
\label{3g}
ds^{2}=-\bigg(1-\frac{2M}{r}\bigg)\,h\,dt^{2}+\bigg(1-\frac{2M}{r}\bigg)^{-1}\!h^{-1}\,dr^{2}+r^{2}\Big(d\theta^{2}+\sin^{2}\theta d\phi^{2}\Big),
\end{align}
where $h=e^{\alpha\,g}$ with $g$ to been determined. The solution obtained in this way is still a Kerr-Schild spacetime ($g_{00}g_{11}=-1$) and the field equations are linear, greatly simplifying any further analyses.

\subsection{Weak energy conditions}
In extremely high-curvature environments, one can expect that the classical energy conditions are generally violated. These conditions, however, are still a good guide to construct physical solutions \cite{Martin-Moruno:2017exc}. Here we require that the tensor $\theta_{\mu\nu}$ fulfills the WEC
\begin{align}
\mathcal{E}\ge 0,                                   \label{WEC1}   \\
\mathcal{E}+\mathcal{P}_{r}\ge 0,                   \label{WEC2}   \\
\mathcal{E}+\mathcal{P}_{\theta}\ge 0.            \label{WEC3}   
\end{align}
After substituting Eqs. \eqref{GD21} and \eqref{Prprime}, the conditions \eqref{WEC1} and \eqref{WEC3} can be respectively rewritten as
\begin{align}
\kappa\,\mathcal{E}\,r^{2}&=-(r-2M)\,h'-h+1\ge 0,                        \label{WEC4}   \\
2(\mathcal{E}+\mathcal{P}_{\theta})&=-r\,\mathcal{E}'\ge 0.              \label{WEC5}   
\end{align}
Obviously, Eq. \eqref{WEC4} is a first-order linear differential inequality for \( h \), and it reduces to the seed Schwarzschild solution \eqref{seedSsolution} when \( \mathcal{E} = 0 \).  
Moreover, any everywhere-regular solution must satisfy the system of equations \eqref{WEC4}–\eqref{WEC5}. This implies that a positive energy density \( \mathcal{E} \) decreases monotonically from \( r = 0 \), i.e., \( \mathcal{E}' \le 0 \). An interesting case we find satisfying all these conditions is
\begin{align}
\label{C1kE}
\kappa\,\mathcal{E}=\frac{\gamma}{\iota^{3}}\big(r+\xi\big)e^{-\frac{r}{\iota}},
\end{align}
where $\gamma$ is a constant, $\xi$ and $\iota$ are positive constants with length dimension. We introduce $\gamma$ so that we can return to the seed vacuum solution \eqref{seedSsolution} when $\gamma\rightarrow0$. While $\xi$ is introduced to meet the WEC requirements. Since \( \mathcal{E} \ge 0 \), meaning that $\gamma\ge 0$. From \eqref{WEC5}, 
$\kappa\,\mathcal{E}'=\gamma(\iota-\xi-r)e^{-\frac{r}{\iota}}/\iota^{4}<0$, we have $1\le\frac{\xi}{\iota}+\frac{r}{\iota}$ for any $r$, implying that $\iota\le\xi$.

\section{Spherically symmetric case}
Substituting \eqref{C1kE} into \eqref{WEC4} and doing some less complicated calculations, we are leaded to
\begin{align}
\label{1h}
h=\frac{r-c_{1}}{r-2M}+\frac{e^{-\frac{r}{\iota}}}{r-2M}\frac{r^{3}\,\gamma+r^{2}\,\gamma\,\big(\xi+3\,\iota\big)+2\,\gamma\,\iota^{2}\,\big(\xi+3\,\iota\big)+2\,r\,\gamma\,\iota\,\big(\xi+3\,\iota\big)}{\iota^{2}},
\end{align}
where $c_{1}$ is also a constant with length dimension. According to Eq. \eqref{1h}, we find that the asymptotically flat metric functions take the following forms
\begin{align}
\label{3g}
e^{A}=e^{-B}=1-\frac{c_{1}}{r}+\frac{\gamma\,e^{-\frac{r}{\iota}}\Big[r^{3}+r^{2}\,\big(\xi+3\,\iota\big)+2\,\iota^{2}\,\big(\xi+3\,\iota\big)+2\,r\,\iota\,\big(\xi+3\,\iota\big)\Big]}{r\,\iota^{2}},
\end{align}
from which the ADM mass is given by $c_{1}=2\mathcal{M}$ and the Schwarzschild solution is recovered for $\gamma\rightarrow 0$. If $r\sim0$, Eqs. (\ref{3g}) approximately equal to 
\begin{align}
\label{4g}
e^{A}=e^{-B}\simeq1+\frac{2\,\gamma\,\xi+6\,\gamma\,\iota-c_{1}}{r}+\frac{\gamma\,r^{2}}{\iota^{2}}+\mathcal{O}(r^{3}).
\end{align}
 The disappearance of the central singularity requires $c_{1}=2\,\gamma\,(\xi+3\,\iota)$, which results in the regularity condition: $\mathcal{M}=\gamma\,(\xi+3\,\iota)$. Finally, in terms of the ADM mass, we can rewrite the metric \eqref{3g} as
\begin{align}
\label{5g}
&e^{A}=e^{-B}=1-\frac{2\,\mathcal{M}}{r}+\frac{e^{-\frac{r}{\iota}}\Big(\gamma\,r^{3}+r^{2}\,\mathcal{M}+2\,\iota^{2}\,\mathcal{M}+2\,r\,\iota\,\mathcal{M}\Big)}{r\,\iota^{2}}                      \nonumber \\
&=1-\frac{2\,\mathcal{M}}{r}+\frac{e^{-\frac{3\,r\,\gamma}{\mathcal{M}-\gamma\,\xi}}\Big[2\,\mathcal{M}^{3}+9\,r^{3}\,\gamma^{3}+\mathcal{M}^{2}\,\gamma\,(6\,r-4\,\xi)+\mathcal{M}\,\gamma^{2}\,(9\,r^{2}-6\,r\,\xi+2\,\xi^{2})\Big]}{r\,(\mathcal{M}-\gamma\,\xi)^{2}}.
\end{align}
After imposing the regularity condition, from Eqs. \eqref{5g} we recover the Minkowski spacetime for $\iota\rightarrow\infty$ (or equivalently $\gamma\rightarrow0$). Now the mass function takes the form
\begin{align}
    \label{m1}
\tilde{m}=\mathcal{M}-\frac{e^{-\frac{r}{\iota}}\Big(\gamma\,r^{3}+r^{2}\,\mathcal{M}+2\,\iota^{2}\,\mathcal{M}+2\,r\,\iota\,\mathcal{M}\Big)}{2\,\iota^{2}},
\end{align}
which, for $r\rightarrow0$, can be simplified as
\begin{align}
\label{m10}
    \tilde{m}\simeq\frac{r^{3}\,(\mathcal{M}-3\,\gamma\,\iota)}{6\,\iota^{3}}+\mathcal{O}(r^{4})=\frac{r^{3}\,\gamma\,\xi}{6\,\iota^{3}}+\mathcal{O}(r^{4}).
\end{align}
This demonstrates that the GD deformation of the seed Schwarzschild metric \eqref{seedSsolution} is a formal procedure. This approach enables the efficient derivation of new black hole solutions with desired physical characteristics, without the need for physical modification of the seed metric.

\begin{figure}[h]
\centering
\includegraphics[width=0.65\textwidth]{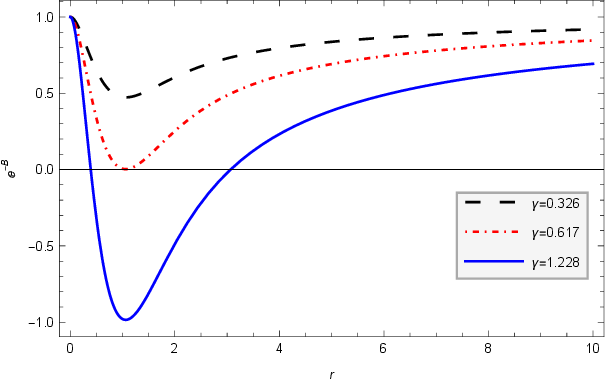}
\caption{Spherically symmetric solutions of the metric function \eqref{5g}: no horizons, on horizon, and two horizons. The extrem case is given by $\gamma^{*}\simeq0.617$, if seting $\xi=0.5$ and $\iota=0.25$.}
\label{t0jie1}
\end{figure}

The possible horizon will be found according to the equation $e^{-B(r_{h})}=0$. Numerical analysis of this equation shows that: there exists an extreme value of $\gamma^{*}$, with no zeros if $\gamma<\gamma^{*}$, exactly one zero for $\gamma=\gamma^{*}$, and two zeros when $\gamma>\gamma^{*}$. As shown in Fig. \ref{t0jie1}, we observe that the metrics display three different cases: no horizon, one horizon, and two horizons which are the Cauchy horizon and the event horizon, respectively.

We find that the curvature scalar and the Ricci squared, respectively, is
\begin{align}
\label{1R}
R=-\frac{e^{-\frac{r}{\iota}}\,\bigg[\gamma\,r^{2}+r\,\big(\mathcal{M}-8\,\gamma\,\iota\big)+4\,\iota\big(-\mathcal{M}+3\,\gamma\,\iota\big)\bigg]}{\iota^{4}},
\end{align}
and
\begin{align}
\label{1RR}
R_{\mu\nu}R^{\mu\nu}=&\frac{e^{-\frac{2\,r}{\iota}}}{2\,\iota^{8}}\,\bigg[\gamma^{2}\,r^{4}+2\,r^{3}\,\gamma\,\big(\mathcal{M}-6\,\gamma\,\iota\big)+8\,\iota^{2}\,\big(\mathcal{M}-3\,\gamma\,\iota\big)^{2}                               \nonumber    \\
&-4\,r\,\iota\big(\mathcal{M}^{2}-11\,\mathcal{M}\,\gamma\,\iota+24\,\gamma^{2}\,\iota^{2}\big)+r^{2}\,\big(\mathcal{M}^{2}-16\,\mathcal{M}\,\gamma\,\iota+52\,\gamma^{2}\,\iota^{2}\big)\bigg].
\end{align}
For the Kretschmann scalar, its complete expression is too involved for displaying, it approximately behaves for $r\rightarrow0$  
\begin{align}
\label{1RRRR}
R_{\mu\nu\rho\sigma}R^{\mu\nu\rho\sigma}\simeq&\frac{8\,\mathcal{M}^{2}}{3\,\iota^{6}}-\frac{16\,\mathcal{M}\,\gamma}{\iota^{5}}+\frac{24\,\gamma^{2}}{\iota^{4}}-\frac{20\,r\,\mathcal{M}^{2}}{3\,\iota^{7}}+\frac{140\,r\,\mathcal{M}\,\gamma}{3\,\iota^{6}}           \nonumber  \\
&-\frac{80\,r\,\gamma^{2}}{\iota^{5}}+\frac{97\,r^{2}\,\mathcal{M}^{2}}{12\,\iota^{8}}-\frac{68\,r^{2}\,\mathcal{M}\,\gamma}{\iota^{7}}+\frac{136\,r^{2}\,\gamma^{2}}{\iota^{6}}+\delta(r^{3}),
\end{align}
from which we can conclude that the solution has no curvature singularity.

\begin{figure}[h]
\centering
\includegraphics[width=0.65\textwidth]{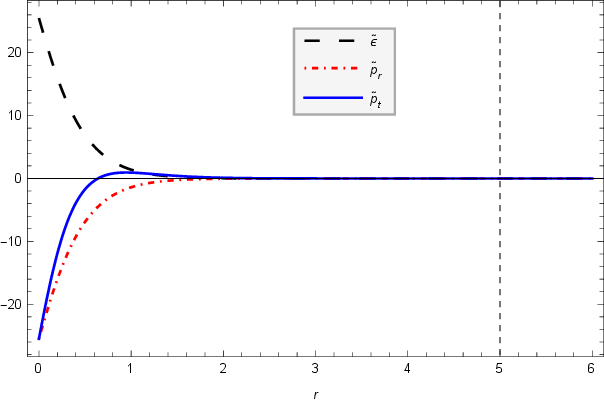}
\caption{The source terms $\{\tilde{\epsilon},\tilde{p}_{r},\tilde{p}_{t}\}\times10$ in the sphericity metric case with $\xi=0.5$, $\iota=0.25$ and $\gamma=2$. The vertical dashed line represents the event horizon $r_{\rm {h}}\sim5$. }
\label{t1EPrP0}
\end{figure}

Finally, the source that generates the metric function \eqref{5g} is characterized by an effective density \eqref{C1kE} and an effective tangential pressure
\begin{align}
\label{P0}
\mathcal{P}_{\theta}=\frac{e^{-\frac{r}{\iota}}\,\gamma\,\big(r^{2}+r\,\xi-3\,r\,\iota-2\,\xi\,\iota\big)}{2\,\kappa\,\iota^{4}},
\end{align}
Combing with Eq. \eqref{C1kE}, we further get
\begin{align}
\label{EP0}
\mathcal{E}+\mathcal{P}_{\theta}=\frac{e^{-\frac{r}{\iota}}\,r\,\gamma\,\big(r+\xi-\iota\big)}{2\,\kappa\,\iota^{4}},
\end{align}
which implies that the WEC holds within the region $r\ge0$, because $\gamma\geq0$ and $\xi>\iota$, as shown in Fig. \ref{t1EPrP0}. Outside the event horizon, the vacuum is approached very rapidly. According to Eq. \eqref{Prprime}, the central pull experienced by the fluid originates from the radial pressure gradient $\mathcal{P}_{r}'\ge0$ is exactly canceled by the gravitational repulsion induced by the pressure anisotropy $\Pi$.

\section{Axially symmetric case}
To construct the rotating form of the metric function \eqref{5g}, we adopt the procedure outlined in \cite{Contreras:2021yxe}.
This corresponds to examining the general Kerr–Schild metric in Boyer–Lindquist coordinates, namely, the Gurses–Gursey metric \cite{Gurses:1975vu}
\begin{align}
\label{6g}
ds^{2}=-\bigg[1-\frac{2\,r\,\tilde{m}(r)}{\rho^{2}}\bigg]dt^{2}-\frac{4 \,a\,r\,\tilde{m}(r)\sin^{2}\theta}{\rho^{2}}\,dtd\phi+\frac{\rho^{2}}{\Delta}\,dr^{2}+\rho^{2}\,d\theta^{2}+\frac{\Sigma\,\sin^{2}\theta}{\rho^{2}}\,d\phi^{2},
\end{align}
with
\begin{align}
\rho^{2}&=r^{2}+a^{2}\cos^{2}\theta,                             \label{rho2}    \\
a&=J/\mathcal{M},                                                \label{a}       \\
\Delta&=r^{2}-2\,r\,\tilde{m}(r)+a^{2},                          \label{Delta}   \\
\Sigma&=(r^{2}+a^{2})^{2}-\Delta\,a^{2}\sin^{2}\theta,           \label{Sigma}
\end{align}
where $J$ is the angular momentum, $\tilde{m}$ given by Eq. \eqref{m1} is the mass function of the reference spherically symmetric metric, and $\mathcal{M}=\tilde{m} (r\rightarrow\infty)$ is the total mass of the system. Obviously, for $\tilde{m}=M$ the metric \eqref{6g} can reduce to the Kerr solution, no need to use the Newman-Janis algorithm.

The energy-momentum tensor $\theta_{\mu\nu}$ generating the metric \eqref{6g} can be re-expressed in terms of orthonormal tetrads as \cite{Gurses:1975vu}
\begin{align}
 \label{tetrad}
\theta^{\mu\nu}=\tilde{\epsilon}\,u^{\mu}\,u^{\nu}+\tilde{p}_{r}\,l^{\mu}\,l^{\nu}+\tilde{p}_{\theta}\,n^{\mu}\,n^{\nu}+\tilde{p}_{\phi}\,m^{\mu}\,m^{\nu},
\end{align}
where
\begin{align}
u^{\mu}&=\frac{\Big(r^{2}+a^{2},0,0,a\Big)}{\sqrt{\Delta\,\rho^{2}}},\qquad                          
l^{\mu}=\frac{\sqrt{\Delta}\Big(0,1,0,0\Big)}{\sqrt{\rho^{2}}},                  \nonumber      \\
n^{\mu}&=\frac{\Big(0,0,1,0\Big)}{\sqrt{\rho^{2}}},\qquad                            
m^{\mu}=-\frac{\Big(a\,\sin^{2}\theta,0,0,1\Big)}{\sqrt{\rho^{2}\,\sin\theta}},                \\
\kappa\,\tilde{\epsilon}&=-\kappa\,\tilde{p}_{r}=\frac{2\,r^{2}}{\rho^{4}}\,\tilde{m}' ,        \\
\kappa\,\tilde{p}_{\theta}&=\kappa\,\tilde{p}_{\phi}=-\frac{r^{2}}{\rho^{2}}\,\tilde{m}''+\frac{2\big(r^{2}-\rho^{2}\big)}{\rho^{4}}\,\tilde{m}'.
\end{align}
The metric \eqref{6g} has two types of singularities: $\rho=0$ or $\Delta=0$. The case of $\rho=0$ corresponds to the ring singularity of the Kerr solution occurring at $\theta=\pi/2$ and $r=0$,which is a physical singularity. When $a=0$, it reduces to a Schwarzschild-like metric.

The curvature scalar of the spacetime geometry \eqref{6g} is given by
\begin{align}
\label{KR}
R=\frac{4\,\tilde{m}'+2\,r\,\tilde{m}''}{\rho^{2}}.
\end{align}
Substituting the mass function \eqref{m1}, it yields
\begin{align}
\label{KerrR}
R=-\frac{e^{-\frac{r}{\iota}}\,r^{2}\,\bigg[r^{2}\,\gamma+r\,\big(\mathcal{M}-8\,\gamma\,\iota\big)+4\,\iota\,\big(-\mathcal{M}+3\,\gamma\,\iota\big)\bigg]}{\iota^{4}\,\rho^{2}}.
\end{align}
It is obvious that for $\theta=\pi/2$ we have $R(r\rightarrow0)=4(\mathcal{M}-3\gamma\iota)/\iota^3$ and $r=0$, meaning that Eq. \eqref{KerrR} is regular. After checking, Ricci squared is completely identical to the regular form presented in Eq. \eqref{1RR}, and Kretschmann scalar is also identical to the regular form presented in Eq. \eqref{1RRRR} for $r\sim0$ and $\theta=\pi/2$. Therefore, we can conclude that the rotating solution has no physical singularities.

\begin{figure}[h]
\centering
\includegraphics[width=0.65\textwidth]{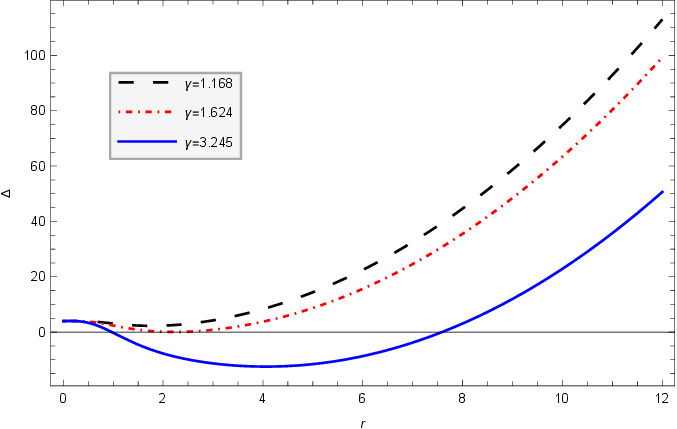}
\caption{Axially symmetric solutions of the metric function \eqref{5g} show three cases: no horizons, one horizon, and two horizons. The extreme BH case is given by $\gamma^{**}\simeq1.624$, if setting $\xi=0.5$, $\iota=0.25$ and $a=2$.}
\label{t0jie2}
\end{figure}

In general, $\Delta=0$ represents a coordinate singularity, implying the existence of a horizon
\begin{align}
\label{0jie2}
\Delta(r_{h})=r_{h}^{2}-2\,r_{h}\,\tilde{m}(r_{h})+a^{2}=0
\end{align}
Note that in general $a\ne0$, Eq. \eqref{0jie2} shows an extreme case: $\gamma=\gamma^{**}$, with one zero. If $\gamma<\gamma^{**}$, there are no zeros. If $\gamma>\gamma^{**}$, there are two zeros, which corresponds to the Cauchy and event horizon, respectively. As shown in Fig. \ref{t0jie2}, we also observe three different cases of the metric: no horizons, one horizon, and two horizons.

\section{conclusions}
A well-known prediction in GR is that the occurrence of singularities is the final result of gravitational collapse. We here introduced a `tensor vacuum’ explicitly described by the expression \eqref{Prprime} which can act like non-collapsing matter to avoid the formation of singularity.

This enables us to construct both static and stationary regular BHs with hairy parameterized by the parameter $\gamma$ which has a clear physical interpretation: for $\gamma\rightarrow0$, it reduces to Minkowski spacetime. While if $\gamma\rightarrow\infty$ and $\gamma\xi\rightarrow0$, the solution \eqref{5g} reduces to Schwarzchild BH, and the solution \eqref{6g} reduces to Kerr BH, respectively. We have also made preliminary limitations on the parameters based on WECs.

In future studies, observational consequences, the stability, time-dependent formation and evaporation for these solutions are worth further investigations.

\begin{acknowledgments}
This study is supported in part by National Natural Science Foundation of China (Grant No. 12333008).
\end{acknowledgments}

\bibliographystyle{elsarticle-num}
\bibliography{Ref}

\begin{thebibliography}{10}
\expandafter\ifx\csname url\endcsname\relax
  \def\url#1{\texttt{#1}}\fi
\expandafter\ifx\csname urlprefix\endcsname\relax\def\urlprefix{URL }\fi
\expandafter\ifx\csname href\endcsname\relax
  \def\href#1#2{#2} \def\path#1{#1}\fi

\bibitem{Martinez:2004nb}
C.~Martinez, R.~Troncoso, J.~Zanelli, {Exact black hole solution with a minimally coupled scalar field}, Phys. Rev. D 70 (2004) 084035.
\newblock \href {http://arxiv.org/abs/hep-th/0406111} {\path{arXiv:hep-th/0406111}}, \href {https://doi.org/10.1103/PhysRevD.70.084035} {\path{doi:10.1103/PhysRevD.70.084035}}.

\bibitem{Sotiriou:2011dz}
T.~P. Sotiriou, V.~Faraoni, {Black holes in scalar-tensor gravity}, Phys. Rev. Lett. 108 (2012) 081103.
\newblock \href {http://arxiv.org/abs/1109.6324} {\path{arXiv:1109.6324}}, \href {https://doi.org/10.1103/PhysRevLett.108.081103} {\path{doi:10.1103/PhysRevLett.108.081103}}.

\bibitem{babichev2014dressing}
E.~Babichev, C.~Charmousis, Dressing a black hole with a time-dependent galileon, Journal of High Energy Physics 2014~(8) (2014) 1--10.

\bibitem{Antoniou:2017acq}
G.~Antoniou, A.~Bakopoulos, P.~Kanti, {Evasion of No-Hair Theorems and Novel Black-Hole Solutions in Gauss-Bonnet Theories}, Phys. Rev. Lett. 120~(13) (2018) 131102.
\newblock \href {http://arxiv.org/abs/1711.03390} {\path{arXiv:1711.03390}}, \href {https://doi.org/10.1103/PhysRevLett.120.131102} {\path{doi:10.1103/PhysRevLett.120.131102}}.

\bibitem{Antoniou:2017hxj}
G.~Antoniou, A.~Bakopoulos, P.~Kanti, {Black-Hole Solutions with Scalar Hair in Einstein-Scalar-Gauss-Bonnet Theories}, Phys. Rev. D 97~(8) (2018) 084037.
\newblock \href {http://arxiv.org/abs/1711.07431} {\path{arXiv:1711.07431}}, \href {https://doi.org/10.1103/PhysRevD.97.084037} {\path{doi:10.1103/PhysRevD.97.084037}}.

\bibitem{Sotiriou:2013qea}
T.~P. Sotiriou, S.-Y. Zhou, {Black hole hair in generalized scalar-tensor gravity}, Phys. Rev. Lett. 112 (2014) 251102.
\newblock \href {http://arxiv.org/abs/1312.3622} {\path{arXiv:1312.3622}}, \href {https://doi.org/10.1103/PhysRevLett.112.251102} {\path{doi:10.1103/PhysRevLett.112.251102}}.

\bibitem{Herdeiro:2015waa}
C.~A.~R. Herdeiro, E.~Radu, {Asymptotically flat black holes with scalar hair: a review}, Int. J. Mod. Phys. D 24~(09) (2015) 1542014.
\newblock \href {http://arxiv.org/abs/1504.08209} {\path{arXiv:1504.08209}}, \href {https://doi.org/10.1142/S0218271815420146} {\path{doi:10.1142/S0218271815420146}}.

\bibitem{Penrose:1969pc}
R.~Penrose, {Gravitational collapse: The role of general relativity}, Riv. Nuovo Cim. 1 (1969) 252--276.
\newblock \href {https://doi.org/10.1023/A:1016578408204} {\path{doi:10.1023/A:1016578408204}}.

\bibitem{hawking2023large}
S.~W. Hawking, G.~F. Ellis, The large scale structure of space-time, Cambridge university press, 2023.

\bibitem{Salazar:1987ap}
I.~H. Salazar, A.~Garcia, J.~Plebanski, {Duality Rotations and Type $D$ Solutions to Einstein Equations With Nonlinear Electromagnetic Sources}, J. Math. Phys. 28 (1987) 2171--2181.
\newblock \href {https://doi.org/10.1063/1.527430} {\path{doi:10.1063/1.527430}}.

\bibitem{Ayon-Beato:1998hmi}
E.~Ayon-Beato, A.~Garcia, {Regular black hole in general relativity coupled to nonlinear electrodynamics}, Phys. Rev. Lett. 80 (1998) 5056--5059.
\newblock \href {http://arxiv.org/abs/gr-qc/9911046} {\path{arXiv:gr-qc/9911046}}, \href {https://doi.org/10.1103/PhysRevLett.80.5056} {\path{doi:10.1103/PhysRevLett.80.5056}}.

\bibitem{Bronnikov:2000vy}
K.~A. Bronnikov, {Regular magnetic black holes and monopoles from nonlinear electrodynamics}, Phys. Rev. D 63 (2001) 044005.
\newblock \href {http://arxiv.org/abs/gr-qc/0006014} {\path{arXiv:gr-qc/0006014}}, \href {https://doi.org/10.1103/PhysRevD.63.044005} {\path{doi:10.1103/PhysRevD.63.044005}}.

\bibitem{Dymnikova:2004zc}
I.~Dymnikova, {Regular electrically charged structures in nonlinear electrodynamics coupled to general relativity}, Class. Quant. Grav. 21 (2004) 4417--4429.
\newblock \href {http://arxiv.org/abs/gr-qc/0407072} {\path{arXiv:gr-qc/0407072}}, \href {https://doi.org/10.1088/0264-9381/21/18/009} {\path{doi:10.1088/0264-9381/21/18/009}}.

\bibitem{Balart:2014cga}
L.~Balart, E.~C. Vagenas, {Regular black holes with a nonlinear electrodynamics source}, Phys. Rev. D 90~(12) (2014) 124045.
\newblock \href {http://arxiv.org/abs/1408.0306} {\path{arXiv:1408.0306}}, \href {https://doi.org/10.1103/PhysRevD.90.124045} {\path{doi:10.1103/PhysRevD.90.124045}}.

\bibitem{Toshmatov:2014nya}
B.~Toshmatov, B.~Ahmedov, A.~Abdujabbarov, Z.~Stuchlik, {Rotating Regular Black Hole Solution}, Phys. Rev. D 89~(10) (2014) 104017.
\newblock \href {http://arxiv.org/abs/1404.6443} {\path{arXiv:1404.6443}}, \href {https://doi.org/10.1103/PhysRevD.89.104017} {\path{doi:10.1103/PhysRevD.89.104017}}.

\bibitem{Fan:2016hvf}
Z.-Y. Fan, X.~Wang, {Construction of Regular Black Holes in General Relativity}, Phys. Rev. D 94~(12) (2016) 124027.
\newblock \href {http://arxiv.org/abs/1610.02636} {\path{arXiv:1610.02636}}, \href {https://doi.org/10.1103/PhysRevD.94.124027} {\path{doi:10.1103/PhysRevD.94.124027}}.

\bibitem{Poisson:1989zz}
E.~Poisson, W.~Israel, {Inner-horizon instability and mass inflation in black holes}, Phys. Rev. Lett. 63 (1989) 1663--1666.
\newblock \href {https://doi.org/10.1103/PhysRevLett.63.1663} {\path{doi:10.1103/PhysRevLett.63.1663}}.

\bibitem{Poisson:1990eh}
E.~Poisson, W.~Israel, {Internal structure of black holes}, Phys. Rev. D 41 (1990) 1796--1809.
\newblock \href {https://doi.org/10.1103/PhysRevD.41.1796} {\path{doi:10.1103/PhysRevD.41.1796}}.

\bibitem{Bonanno:2020fgp}
A.~Bonanno, A.-P. Khosravi, F.~Saueressig, {Regular black holes with stable cores}, Phys. Rev. D 103~(12) (2021) 124027.
\newblock \href {http://arxiv.org/abs/2010.04226} {\path{arXiv:2010.04226}}, \href {https://doi.org/10.1103/PhysRevD.103.124027} {\path{doi:10.1103/PhysRevD.103.124027}}.

\bibitem{carballo2022regular}
R.~Carballo-Rubio, F.~Di~Filippo, S.~Liberati, C.~Pacilio, M.~Visser, Regular black holes without mass inflation instability, Journal of High Energy Physics 2022~(9) (2022) 1--14.

\bibitem{Franzin:2022wai}
E.~Franzin, S.~Liberati, J.~Mazza, V.~Vellucci, {Stable rotating regular black holes}, Phys. Rev. D 106~(10) (2022) 104060.
\newblock \href {http://arxiv.org/abs/2207.08864} {\path{arXiv:2207.08864}}, \href {https://doi.org/10.1103/PhysRevD.106.104060} {\path{doi:10.1103/PhysRevD.106.104060}}.

\bibitem{Bonanno:2022jjp}
A.~Bonanno, A.-P. Khosravi, F.~Saueressig, {Regular evaporating black holes with stable cores}, Phys. Rev. D 107~(2) (2023) 024005.
\newblock \href {http://arxiv.org/abs/2209.10612} {\path{arXiv:2209.10612}}, \href {https://doi.org/10.1103/PhysRevD.107.024005} {\path{doi:10.1103/PhysRevD.107.024005}}.

\bibitem{Ovalle:2023ref}
J.~Ovalle, R.~Casadio, A.~Giusti, {Regular hairy black holes through Minkowski deformation}, Phys. Lett. B 844 (2023) 138085.
\newblock \href {http://arxiv.org/abs/2304.03263} {\path{arXiv:2304.03263}}, \href {https://doi.org/10.1016/j.physletb.2023.138085} {\path{doi:10.1016/j.physletb.2023.138085}}.

\bibitem{Ovalle:2017fgl}
J.~Ovalle, {Decoupling gravitational sources in general relativity: from perfect to anisotropic fluids}, Phys. Rev. D 95~(10) (2017) 104019.
\newblock \href {http://arxiv.org/abs/1704.05899} {\path{arXiv:1704.05899}}, \href {https://doi.org/10.1103/PhysRevD.95.104019} {\path{doi:10.1103/PhysRevD.95.104019}}.

\bibitem{Ovalle:2018gic}
J.~Ovalle, {Decoupling gravitational sources in general relativity: The extended case}, Phys. Lett. B 788 (2019) 213--218.
\newblock \href {http://arxiv.org/abs/1812.03000} {\path{arXiv:1812.03000}}, \href {https://doi.org/10.1016/j.physletb.2018.11.029} {\path{doi:10.1016/j.physletb.2018.11.029}}.

\bibitem{Ovalle:2018umz}
J.~Ovalle, R.~Casadio, R.~d. Rocha, A.~Sotomayor, Z.~Stuchlik, {Black holes by gravitational decoupling}, Eur. Phys. J. C 78~(11) (2018) 960.
\newblock \href {http://arxiv.org/abs/1804.03468} {\path{arXiv:1804.03468}}, \href {https://doi.org/10.1140/epjc/s10052-018-6450-4} {\path{doi:10.1140/epjc/s10052-018-6450-4}}.

\bibitem{Ovalle:2020kpd}
J.~Ovalle, R.~Casadio, E.~Contreras, A.~Sotomayor, {Hairy black holes by gravitational decoupling}, Phys. Dark Univ. 31 (2021) 100744.
\newblock \href {http://arxiv.org/abs/2006.06735} {\path{arXiv:2006.06735}}, \href {https://doi.org/10.1016/j.dark.2020.100744} {\path{doi:10.1016/j.dark.2020.100744}}.

\bibitem{Contreras:2021yxe}
E.~Contreras, J.~Ovalle, R.~Casadio, {Gravitational decoupling for axially symmetric systems and rotating black holes}, Phys. Rev. D 103~(4) (2021) 044020.
\newblock \href {http://arxiv.org/abs/2101.08569} {\path{arXiv:2101.08569}}, \href {https://doi.org/10.1103/PhysRevD.103.044020} {\path{doi:10.1103/PhysRevD.103.044020}}.

\bibitem{Islam:2021dyk}
S.~U. Islam, S.~G. Ghosh, {Strong field gravitational lensing by hairy Kerr black holes}, Phys. Rev. D 103~(12) (2021) 124052.
\newblock \href {http://arxiv.org/abs/2102.08289} {\path{arXiv:2102.08289}}, \href {https://doi.org/10.1103/PhysRevD.103.124052} {\path{doi:10.1103/PhysRevD.103.124052}}.

\bibitem{daRocha:2020gee}
R.~da~Rocha, A.~A. Tomaz, {MGD-decoupled black holes, anisotropic fluids and holographic entanglement entropy}, Eur. Phys. J. C 80~(9) (2020) 857.
\newblock \href {http://arxiv.org/abs/2005.02980} {\path{arXiv:2005.02980}}, \href {https://doi.org/10.1140/epjc/s10052-020-8414-8} {\path{doi:10.1140/epjc/s10052-020-8414-8}}.

\bibitem{Ovalle:2021jzf}
J.~Ovalle, E.~Contreras, Z.~Stuchlik, {Kerr{\textendash}de Sitter black hole revisited}, Phys. Rev. D 103~(8) (2021) 084016.
\newblock \href {http://arxiv.org/abs/2104.06359} {\path{arXiv:2104.06359}}, \href {https://doi.org/10.1103/PhysRevD.103.084016} {\path{doi:10.1103/PhysRevD.103.084016}}.

\bibitem{afrin2021parameter}
M.~Afrin, R.~Kumar, S.~G. Ghosh, Parameter estimation of hairy kerr black holes from its shadow and constraints from m87, Monthly Notices of the Royal Astronomical Society 504~(4) (2021) 5927--5940.

\bibitem{Ramos:2021jta}
A.~Ramos, C.~Arias, R.~Avalos, E.~Contreras, {Geodesic motion around hairy black holes}, Annals Phys. 431 (2021) 168557.
\newblock \href {http://arxiv.org/abs/2107.01146} {\path{arXiv:2107.01146}}, \href {https://doi.org/10.1016/j.aop.2021.168557} {\path{doi:10.1016/j.aop.2021.168557}}.

\bibitem{Meert:2021khi}
P.~Meert, R.~da~Rocha, {Gravitational decoupling, hairy black holes and conformal anomalies}, Eur. Phys. J. C 82~(2) (2022) 175.
\newblock \href {http://arxiv.org/abs/2109.06289} {\path{arXiv:2109.06289}}, \href {https://doi.org/10.1140/epjc/s10052-022-10121-6} {\path{doi:10.1140/epjc/s10052-022-10121-6}}.

\bibitem{Mahapatra:2022xea}
S.~Mahapatra, I.~Banerjee, {Rotating hairy black holes and thermodynamics from gravitational decoupling}, Phys. Dark Univ. 39 (2023) 101172.
\newblock \href {http://arxiv.org/abs/2208.05796} {\path{arXiv:2208.05796}}, \href {https://doi.org/10.1016/j.dark.2023.101172} {\path{doi:10.1016/j.dark.2023.101172}}.

\bibitem{Cavalcanti:2022cga}
R.~T. Cavalcanti, R.~C. de~Paiva, R.~da~Rocha, {Echoes of the gravitational decoupling: scalar perturbations and quasinormal modes of hairy black holes}, Eur. Phys. J. Plus 137~(10) (2022) 1185.
\newblock \href {http://arxiv.org/abs/2203.08740} {\path{arXiv:2203.08740}}, \href {https://doi.org/10.1140/epjp/s13360-022-03407-x} {\path{doi:10.1140/epjp/s13360-022-03407-x}}.

\bibitem{Omwoyo:2021uah}
E.~Omwoyo, H.~Belich, J.~C. Fabris, H.~Velten, {Remarks on the black hole shadows in Kerr-de Sitter space times}, Eur. Phys. J. C 82~(5) (2022) 395.
\newblock \href {http://arxiv.org/abs/2112.14124} {\path{arXiv:2112.14124}}, \href {https://doi.org/10.1140/epjc/s10052-022-10361-6} {\path{doi:10.1140/epjc/s10052-022-10361-6}}.

\bibitem{Avalos:2023jeh}
R.~Avalos, E.~Contreras, {Quasi normal modes of hairy black holes at higher-order WKB approach}, Eur. Phys. J. C 83~(2) (2023) 155.
\newblock \href {http://arxiv.org/abs/2302.09148} {\path{arXiv:2302.09148}}, \href {https://doi.org/10.1140/epjc/s10052-023-11288-2} {\path{doi:10.1140/epjc/s10052-023-11288-2}}.

\bibitem{Avalos:2023ywb}
R.~Avalos, P.~Bargue{\~n}o, E.~Contreras, {A Static and Spherically Symmetric Hairy Black Hole in the Framework of the Gravitational Decoupling}, Fortsch. Phys. 71~(4-5) (2023) 2200171.
\newblock \href {http://arxiv.org/abs/2303.04119} {\path{arXiv:2303.04119}}, \href {https://doi.org/10.1002/prop.202200171} {\path{doi:10.1002/prop.202200171}}.

\bibitem{Ovalle:2020fuo}
J.~Ovalle, R.~Casadio, {Beyond Einstein Gravity}: {The Minimal Geometric Deformation Approach in the Brane-World}, SpringerBriefs in Physics, Springer, 2020.
\newblock \href {https://doi.org/10.1007/978-3-030-39493-6} {\path{doi:10.1007/978-3-030-39493-6}}.

\bibitem{Martin-Moruno:2017exc}
P.~Martin-Moruno, M.~Visser, {Classical and semi-classical energy conditions}, Fundam. Theor. Phys. 189 (2017) 193--213.
\newblock \href {http://arxiv.org/abs/1702.05915} {\path{arXiv:1702.05915}}, \href {https://doi.org/10.1007/978-3-319-55182-1_9} {\path{doi:10.1007/978-3-319-55182-1_9}}.

\bibitem{Gurses:1975vu}
M.~Gurses, F.~Gursey, {Lorentz Covariant Treatment of the Kerr-Schild Metric}, J. Math. Phys. 16 (1975) 2385.
\newblock \href {https://doi.org/10.1063/1.522480} {\path{doi:10.1063/1.522480}}.

\end{thebibliography}

\end{document}